\documentclass[3p,twocolumn]{elsarticle}
\usepackage{ecrc}

\volume{00}

\firstpage{1}
\journalname{Nuclear Physics B Proceedings Supplement}
\runauth{Tamas Almos Vami}
\jid{nuphbp}
\jnltitlelogo{Nuclear Physics B Proceedings Supplement}

\usepackage{amssymb}
\usepackage{lineno}

\begin{document}
\begin{frontmatter}

\title{Calibration of the CMS Pixel Detector at the Large Hadron Collider}

\author[1]{Tamas Almos Vami}
\author{on behalf of the CMS Collaboration}
\address[1]{Wigner Research Centre for Physics, Budapest}

\begin{abstract}
The Compact Muon Solenoid (CMS) detector is one of two general-purpose detectors that
 reconstruct the products of high energy particle interactions at the Large
  Hadron Collider (LHC) at CERN. The silicon pixel detector is the innermost component of the CMS tracking system. It determines the trajectories of
   charged particles originating from the interaction region in three points with high resolution enabling precise momentum and impact parameter measurements in the tracker. The pixel detector is exposed to intense 
ionizing radiation generated by particle collisions in the LHC. This 
irradiation could result in temporary or permanent malfunctions of the sensors and could decrease the efficiency of the detector. We have developed procedures in order to correct for these effects. In 
this paper, we present the types of malfunctions and the offline calibration procedures. We will also show the efficiency and the resolution of the detector in  2012.
\end{abstract}


\end{frontmatter}

\section{Introduction}
\label{sec::Introduction}
The CMS detector is a general-purpose detector at the LHC at CERN~\cite{CMS_exp}. The task of the CMS Tracker \cite{Tracker1} is to measure the trajectories of charged particles with high accuracy and reconstruction efficiency \cite{Tracker3}.

The pixel detector is constructed in hybrid pixel technology. The barrel
pixel detector (BPIX) consists of three cylindric layers of pixel modules oriented coaxially with respect to the beamline and centred around the
collision point. The forward pixel detector (FPIX) includes a pair of disks per side composed of pixel modules, which are situated orthogonally to the beamline on each side of BPIX.

The sensors of the pixel detector are segmented into a total of 66 million pixels, each with the size of 100\,$\mathrm{{\mu}m}$ $\times$ 150\,$\mathrm{{\mu}m}$. A 52\,$\times$\,80 pixel array is processed by one read-out chip.

\section{Track reconstruction}

The tracks of electrically charged particles are determined via a multi-stage process. 

In the first step, traversing particles are detected through hit pixels which have deposited charge above a certain threshold. Subsequently, these pixels are combined into clusters. These clusters are called hits after they are characterised by their charge and their 2D position. Hit reconstruction is based on the use of pre-computed, projected cluster shapes~\cite{Swartz}.

The hits are combined into tracks using a tracking algorithm. The trajectory building algorithm extrapolates tracks using hits estimated from former hits along
the track. The algorithm stops if no hit is found on two consecutive layers on the projected trajectory.

\section{Detector calibration}
\label{sec::DetectorCalibration}

\subsection{Bad component database}
\label{sec::Bad component database}
The bad component database contains the list of permanently or temporarily bad modules. It enables the trajectory building algorithm to skip a tracking layer on which a hit is expected to be missing due to a faulty module.

\begin{figure}[!h]
\centering
\includegraphics[scale=0.15]{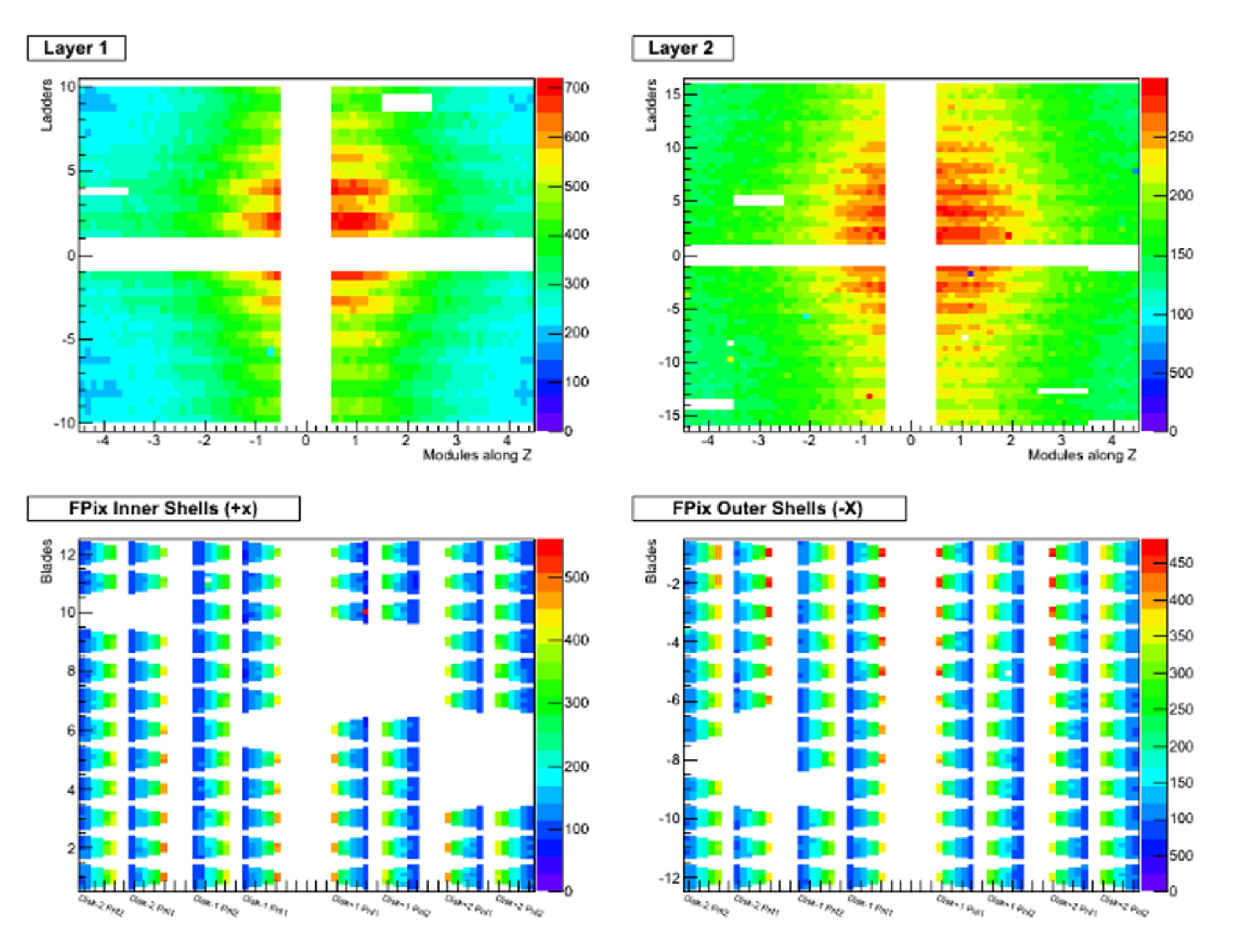} 
\caption{Occupancy map of the sensors in 2012. Layer 1 and Layer 2 in BPIX (upper plots) and the two disks in FPIX (lower plots). Color code shows the occupancy of the sensors. The coordinates x,y,z refer to the local coordinate system of the CMS \citep{CMS_exp}.}
\label{fig::bad component}
\end{figure}

Bad modules are determined by measuring their occupancy. A map is created, in which the faulty sensors are identified as those with significantly lower occupancy compared to the average of the modules surrounding them. Figure~\ref{fig::bad component} shows the map of the sensors where the white areas correspond to the bad modules. White horizontal and vertical stripes in the middle of the maps are due to the fact that the coordinate 0 does not designate any ladders or modules.

\subsection{Single Event Upsets}
Temporary faults in modules are caused by ionising radiation that can flip the memory state of the read-out chip. 
This effect is called Single Event Upset (SEU).

The bad component list is updated every 23 seconds in order to reflect not only permanently bad modules, but also temporarily faulty ones which undergo a SEU.
The SEUs are fixed by reprogramming the read-out chips. Using an automatic online monitoring and recovery system we were able to recover a detector efficiency of 0.05\% per hour. 

\section{Efficiency}
\label{sec::Efficiency}

\begin{figure}[!h]
\centering
\includegraphics[scale=0.25]{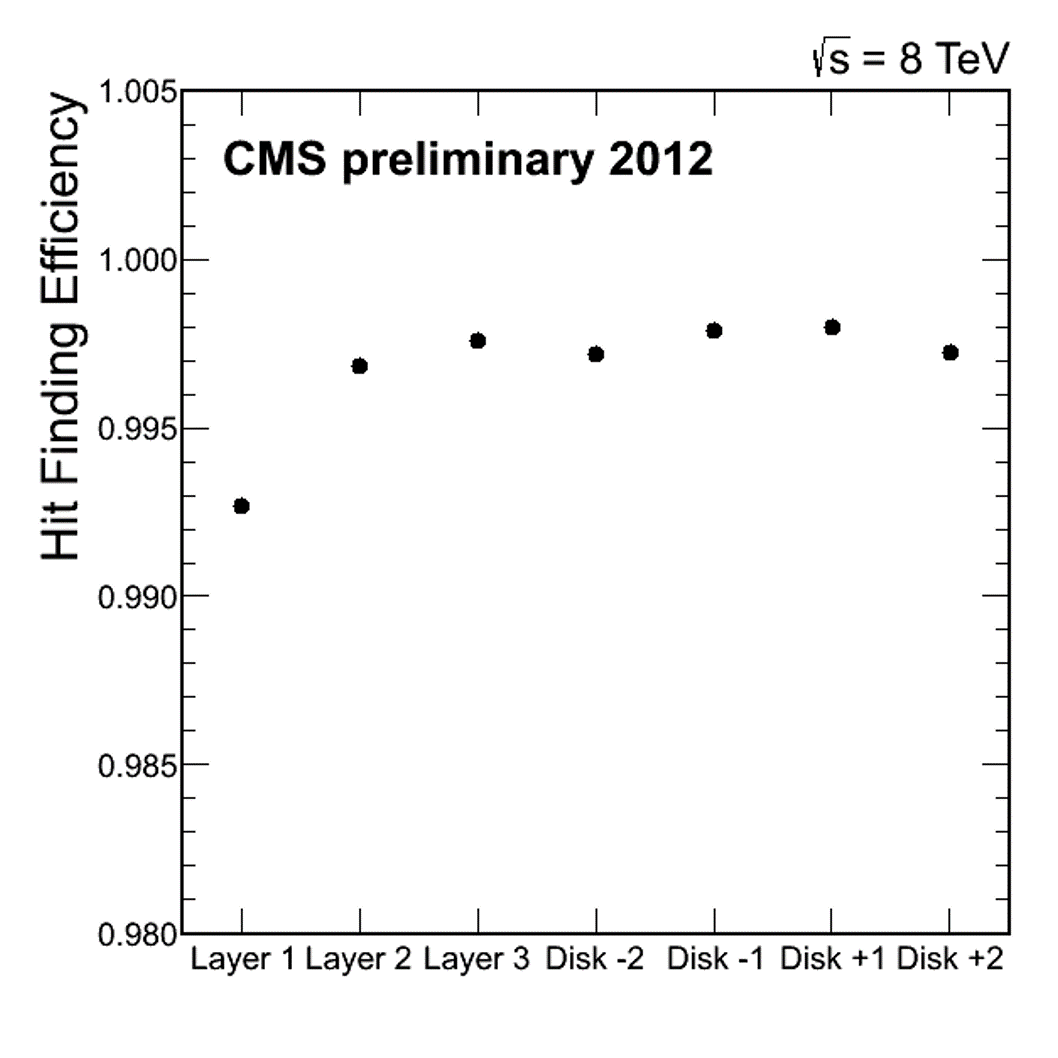} 
\caption{Average hit finding efficiency of the pixel detector in 2012.}
\label{fig::Eff}
\end{figure}

Hit efficiency is defined as the detected fraction of all expected clusters in a fiducial region. After excluding components which are listed in the bad component database the efficiency was measured to be above 99.5\%, except for the first layer (Fig.~\ref{fig::Eff}) 
The efficiency on the first layer is lower, because of losing hit information during the trigger latency period due to buffer overflow in the read-out chips.

\section{Resolution and cluster size}
\label{sec::Resolution and cluster size}
\subsection{Hit resolution}
\begin{figure}
\centering
\includegraphics[scale=0.25]{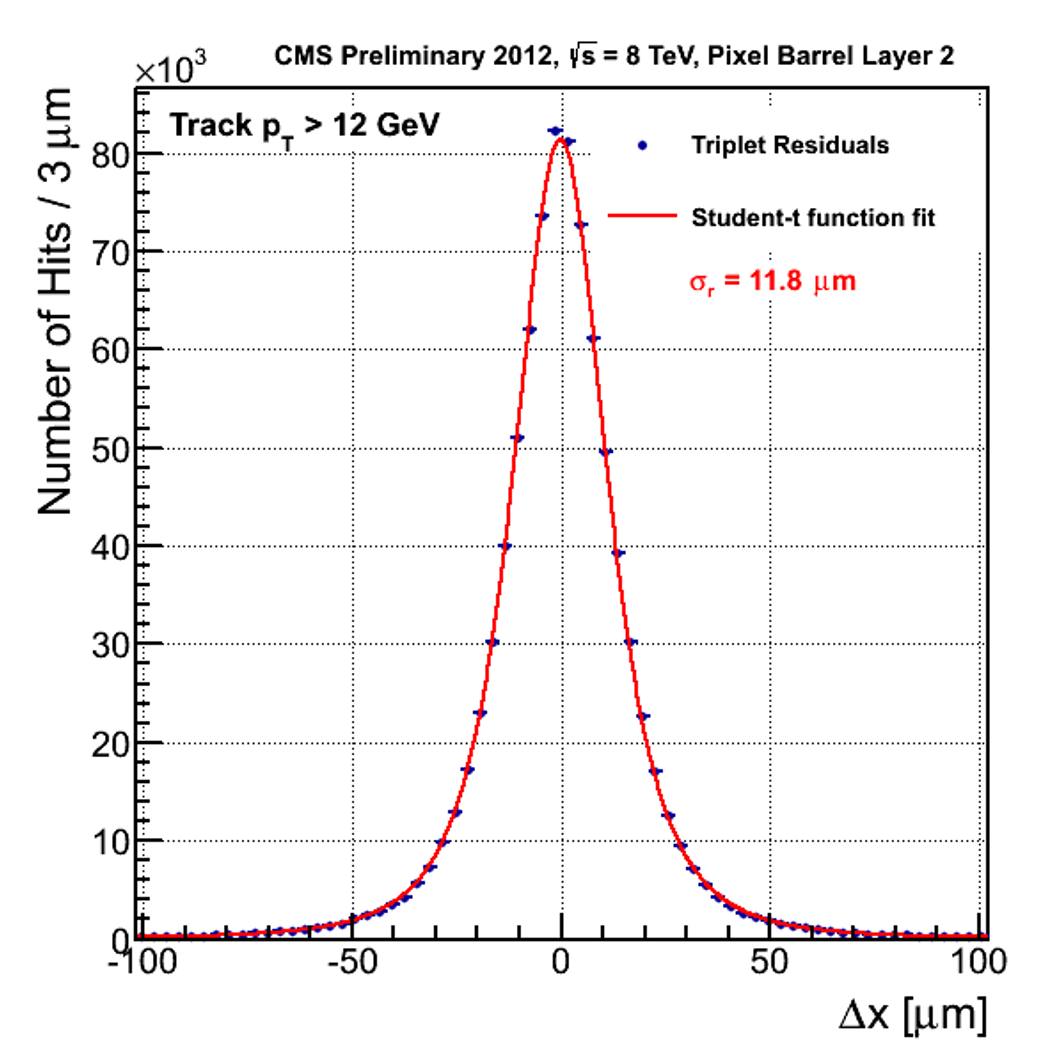} 
\caption{The residual difference between the hit position and the
interpolated track, and the student-t function fit in Layer 2 in the transverse plane.}
\label{fig::BPIX_res}
\end{figure}

The resolution of the cluster position measurement determines the accuracy of the track reconstruction. The resolution is measured from the hit residuals obtained in the track fitting. The hits along the tracks are re-fitted without the hit in the examined layer. 
The distribution of the differences between the hits  and the
interpolated track positions are shown in Fig.~\ref{fig::BPIX_res}. A student-t function is fitted to the distribution. Resolution is derived from the width of the function using a method described in Ref. \cite{Burgmeier}. 

Hits can be reconstructed with an accuracy of about 10\,$\mathrm{{\mu}m}$ in Layer 2 in the azimuthal direction.

\subsection{Lorentz angle measurement}
The charge carriers, induced by the traversing particles inside the silicon bulk, are deflected by the Lorentz force due to the 3.8 T magnetic field of the CMS magnet.
This drift is characterised by the angle of deflection, which is known as the Lorentz angle. It is defined as the angle between the direction perpendicular to the surface of the sensor and the path of the electrons inside the sensor. Detailed information can be found in Ref.~\cite{Henrich}.

\begin{figure}[!h]
\centering
\includegraphics[scale=0.25]{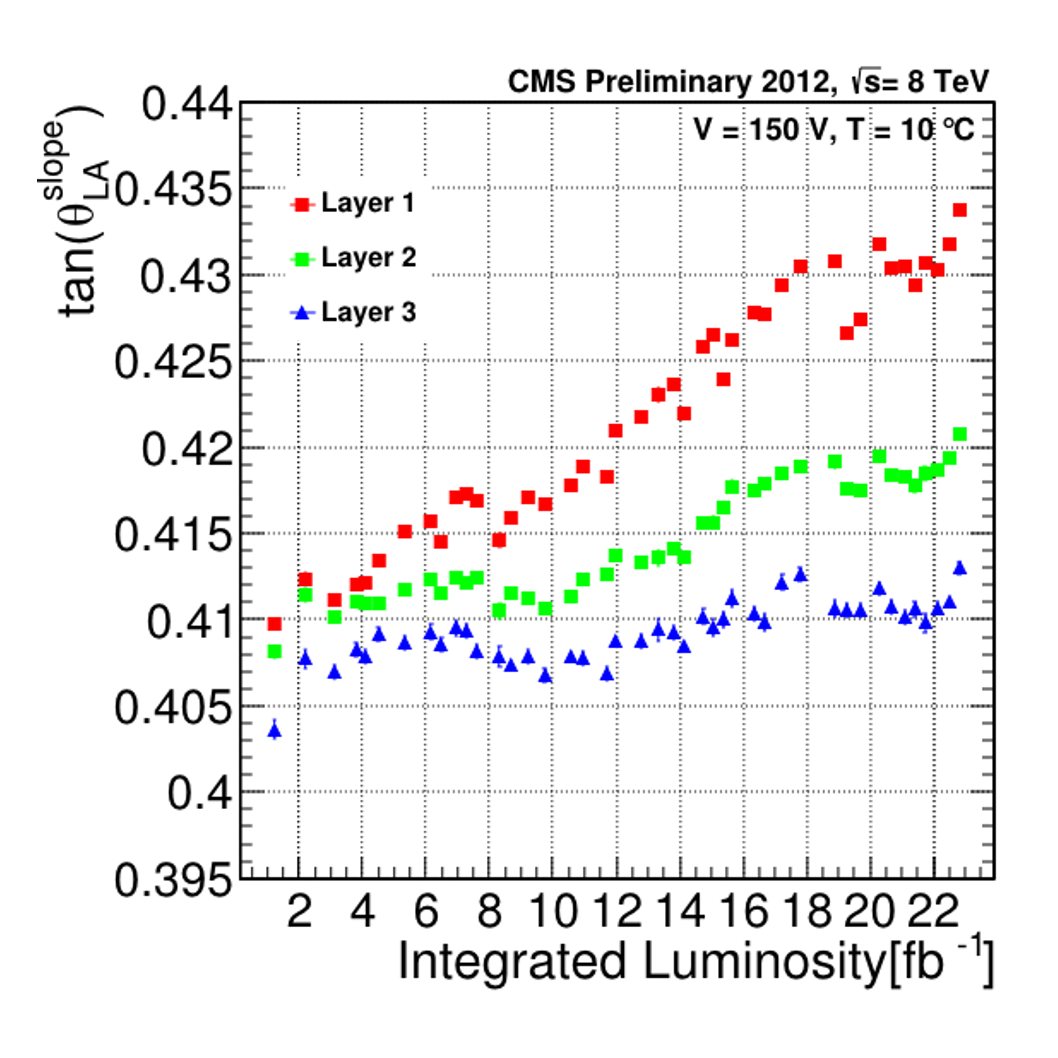} 
\caption{Lorentz angle as a function of integrated luminosity in 2012.}
\label{fig::LA}
\end{figure}

The Lorentz angle was determined as a function of integrated luminosity for 2012 data (Fig.~\ref{fig::LA}). It increases with integrated luminosity as a result of irradiation effects. 

\subsection{Cluster size}
\begin{figure}
\centering
\includegraphics[scale=0.25]{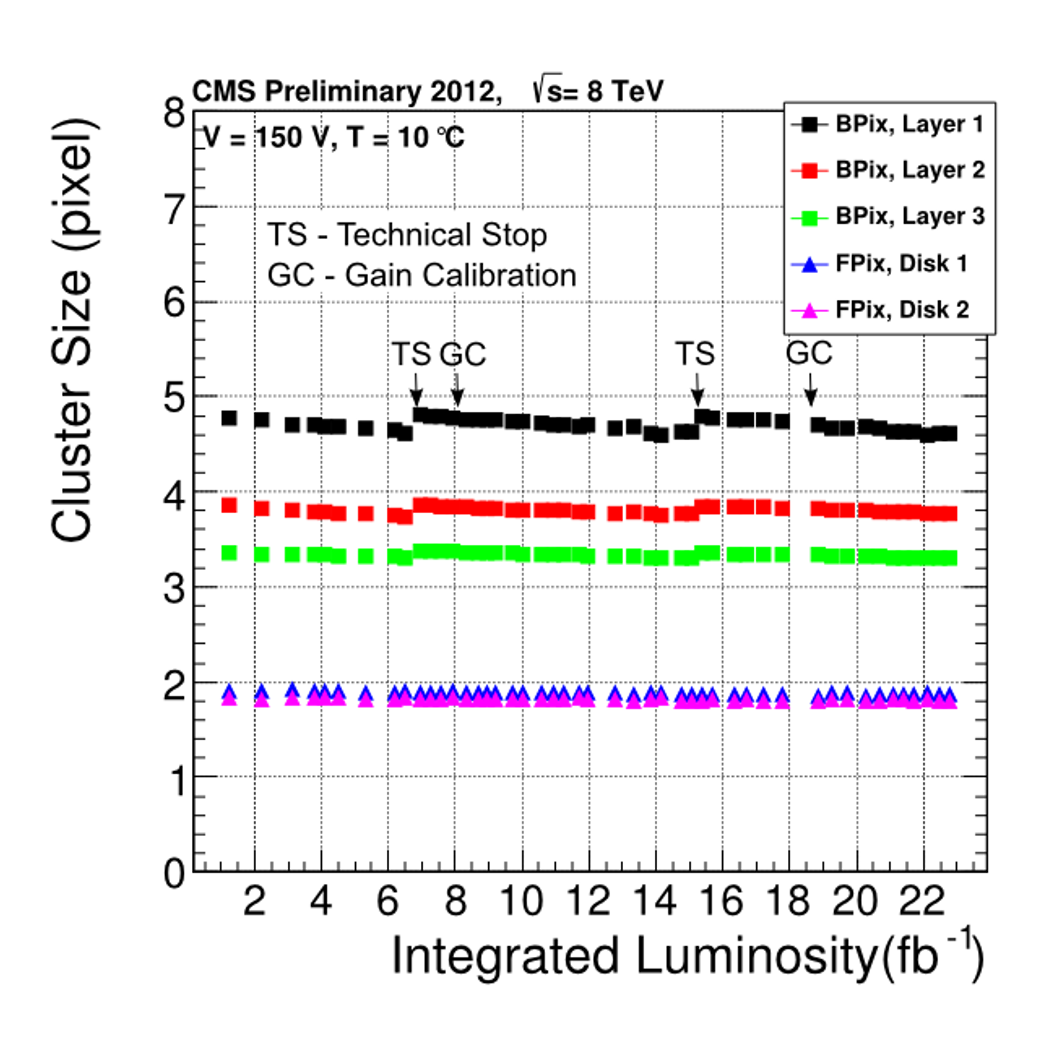} 
\caption{Cluster size as a function of integrated luminosity.}
\label{fig::cluster_size}
\end{figure}

The average number of pixels in a cluster was also monitored as a function of integrated luminosity (Fig.~\ref{fig::cluster_size}). 

The sudden enhancements at $6.5 \mathrm{~fb^{-1}}$ and $15 \mathrm{~fb^{-1}}$ correspond to threshold readjustments during LHC technical stops.

\section{Conclusion}
\label{sec::Conclusion}
The pixel detector worked reliably with high efficiency in 2012. Its efficiency was preserved by the online monitoring and recovery system. Its resolution is in the order of 10 $\mathrm{{\mu}m}$ in the transverse plane. 


\nocite{*}
\bibliography{ref.bib}
\bibliographystyle{elsarticle-num}

\end{document}